\documentclass[twocolumn,showpacs,preprintnumbers,amsmath,amssymb]{revtex4}
\usepackage{graphicx}% Include figure files
\usepackage{dcolumn}% Align table columns on decimal point
\usepackage{bm}% bold math

\begin{document}

%\preprint{APS/123-QED}

\title{Calibrating dipolar interaction in an atomic condensate}
\author{S. Yi$^{1,\dag}$ and L. You$^{1,2}$}

\affiliation{$^1$School of Physics, Georgia Institute of
Technology, Atlanta, GA 30332-0430, USA}
\affiliation{$^2$Interdisciplinary Center of Theoretical Studies
and Institute of Theoretical Physics, CAS, Beijing 10080, China}

\begin{abstract}
We revisit the topic of a dipolar condensate with the recently
derived more rigorous pseudo-potential for dipole-dipole
interaction [A. Derevianko, Phys. Rev. A {\bf 67}, 033607 (2003)].
Based on the highly successful variational technique, we find that
all dipolar effects estimated before (using the bare dipole-dipole
interaction) become significantly larger, i.e. are amplified by
the new velocity-dependent pseudo-potential, especially in the
limit of large or small trap aspect ratios. This result points to
a promising prospect for detecting dipolar effects inside an
atomic condensate.
\end{abstract}

\date{\today}

\maketitle

Interactions make life interesting. To a large
degree, they determine both the kinematic and dynamic properties
of a physical system. In recent years, atomic quantum gases have
become testing grounds for investigating interaction effects.
At typical temperatures for these quantum gases,
the dominant interaction, the binary atomic elastic
collision, is isotropic and thus characterized by the s-wave scattering
length $a_{\rm ss}$. The manipulation of its strength from
strong to weak, and its character from attractive to negative
with a Feshbach resonance, has
become one of the continuing highlights.
At a more detailed level, however, atoms are composite particles, e.g.
possessing magnetic dipole moments, from the electron
and the nuclear spin. The resulting dipolar interaction between atoms,
is anisotropic, and constitutes an exciting new development.
While much weaker as compared to typical isotropic s-wave
interaction, its experimental detection
is only a matter of time, considering the rapid pace
of advances in this active field.

For a condensate of $N$ atoms interacting via a potential
$V(\vec r-\vec r\,')$, the total energy functional is
\begin{eqnarray}
E[\psi^*(\vec r),\psi(\vec r)]=\int d\vec r\psi^*(\vec
r)\left[-\frac{\hbar^2\nabla^2}{2M}+V_{\rm ext}(\vec
r)\right]\psi(\vec r)\nonumber\\
+\frac{1}{2}\int d\vec r d\vec r\,'\psi^*(\vec r)\psi^*(\vec
r\,')V(\vec R)\psi(\vec r\,')\psi(\vec r), \label{ham}
\end{eqnarray}
where $\psi(\vec r)$ is the condensate wave function and $\vec
R=\vec r-\vec r\,'$. $V_{\rm ext}(\vec
r)=M\omega_\rho^2(x^2+y^2+\lambda^2z^2)/2$ is the trap potential,
assumed harmonic and of axial symmetry with
radial (axial) trap frequency $\omega_\rho$ ($\omega_z=\lambda\omega_\rho$).

The real (bare) potential $V(\vec R)$ in the interaction energy
[the 2nd line of Eq. (\ref{ham})] is usually replaced by a
pseudo-potential $\hat{\cal V}$, which for an isotropic short
ranged interaction takes the contact form
\begin{eqnarray}
\hat{\cal V}(\vec R)=g\delta(\vec R),
\label{asc}
\end{eqnarray}
with $g=4\pi\hbar^2a_{\rm ss}/M$. $a_{\rm ss}$ is the s wave scattering length
of $V(\vec R)$. To date, this pseudo-potential approach
has proven remarkably effective for most studies. %\cite{one}.

In this Letter we revisit the topic of a
condensate of polarized atoms (along $\hat z$)
including dipolar interaction
\begin{eqnarray}
V_{\rm DD}(\vec R)=\frac{C_3}{R^3}P_2(\cos\theta_R),
\label{intdd}
\end{eqnarray}
where $\theta_R$ is the polar angle of $\vec R$, and $P_2(.)$ is
the 2nd order Legendre polynomial. This problem is important
because the non-spherically symmetric interaction Eq. (\ref{intdd})
leads to interesting low energy collisions
due to the presence of
both the `short-' and `long-' range characters \cite{mircea}.
We have previously suggested a pseudo-potential
\begin{eqnarray}
\hat{\cal V}_{\rm DD}^{\rm Born}(\vec R)
=g\delta(\vec R)+\frac{C_3}{R^3}P_2(\cos\theta_R),
\label{redd}
\end{eqnarray}
by matching its first
Born amplitudes to the complete scattering
amplitudes of $V(\vec R)$ \cite{yi1} and confirmed
its validity, provided the dipole moment is not much larger than
a Bohr magneton ($\mu_B$) and the collision is away from any
shape resonances \cite{mircea}.

Atomic dipolar condensate with interaction
(\ref{redd}) has been studied by many groups.
A lot has been learnt about its ground state
and the associated stability \cite{yi1,yi2,goral1,santos},
collective excitations \cite{yi3,goral2,santos2,dell}, free
expansion dynamics \cite{yi4,giovanazzi1}, and the potential
existence of several exotic phases in an optical
lattice \cite{goral3}. Recently, it was discovered that the ground state
density profile remains an inverted parabola in the Thomas-Fermi limit \cite{dtf}.

Following Huang and Yang \cite{huang}
for spherically symmetric potentials, Derevianko
recently proposed a more rigorous pseudo-potential
$\hat{\cal V}_{\rm ansi}$
applicable to anisotropic interactions
including regions near collision resonances \cite{andrei1}.
Due to its dependence on the relative momentum,
the interaction energy of $\hat{\cal V}_{\rm ansi}$
is most conveniently expressed in
momentum representation as \cite{andrei1,andrei2}
\begin{eqnarray}
E_{\rm int}&=&4\int d\vec k_1d\vec kd\vec k'\phi^*(\vec
k_1)\phi^*(\vec k_1-2\vec k)v(\vec k,\vec k')\nonumber\\
&&\times\phi(\vec k_1+\vec k'-\vec k)\phi(\vec k_1-\vec k'-\vec
k),\label{int}
\end{eqnarray}
where $\phi(\vec k)$ is the Fourier transform of $\psi(\vec r)$.
$\vec k$ and $\vec k\,'$ are, respectively, one half the pre-
and post-collision relative momenta of the colliding pair.
The momentum representation of $\hat{\cal V}_{\rm ansi}$,
denoted by
$v(\vec k,\vec k')$ takes the form \cite{andrei1}
\begin{eqnarray}
v(\vec k,\vec k')
&=&\frac{\hbar^2}{2\pi^2M}\left[a_{\rm ss}
-a_{\rm sd}{\cal F}_{DD}(\vec k,\vec k')\right],
\label{psedd}
\end{eqnarray}
with $a_{\rm sd}$ the generalized scattering length due to the coupling
of the $s$ $(l=0)$ and $d$ $(l=2)$ partial wave channels.
Both $a_{\rm ss}$ and
$a_{\rm sd}$ are obtained from the zero energy T-matrix elements
of a coupled multi-channel scattering calculation \cite{mircea}.
In the low energy limit \cite{andrei1},
$$
{\cal F}_{DD}(\vec k,\vec k')=\sqrt{5}P_2(\cos
\theta_{k'})+\frac{3}{\sqrt{5}}
\left(\frac{k}{k'}\right)^2P_2(\cos\theta_{k}),
$$
where the 1st term is momentum-independent, essentially
corresponds to the bare dipolar potential Eq. (\ref{intdd});
the 2nd term, on the other hand, depends on the momenta. It arises
from the rigorous construction of the pseudo-potential. The main
purpose of this study is to calibrate how it modifies the
properties of a dipolar condensate. To our surprise, we find
previous studies without the momentum-dependent
2nd term have severely under-estimated the strength of the
dipolar interaction.

Although $\hat{\cal V}_{\rm ansi}$ is non-Hermitian, a variational study
can nevertheless be performed within the mean field theory.
$\hat{\cal V}_{\rm ansi}$ is constructed by matching its (two-body)
scattering solution in the asymptotical ($R$ large) limit
to that of the real potential $V(\vec R)$ \cite{andrei1}.
It is non-Hermitian because its scattering solution differs from
the real one in the short-range.
Within the mean field approximation, however, all condensed atoms share
the same spatial orbital, a smoothed
or ``coarse grained" condensate wave function
(valid over length scales much larger than the range of atomic interactions).
Thus the mean field theory is limited to a subspace of
nonsingular functions of the complete Hilbert space,
where the pseudo-potential $\hat{\cal V}_{\rm ansi}$ is Hermitian
and leads to an unitary time evolution.
As expected, with a Gaussian ansatz
\begin{eqnarray}
\psi(\vec
r)=\frac{N^{1/2}}{\pi^{3/4}(w_\rho^2w_z)^{1/2}}\exp\left[-\frac{1}{2}\left(
\frac{x^2+y^2}{w_\rho^2}+\frac{z^2}{w_z^2}\right)\right],
\label{trial}
\end{eqnarray}
of variation parameters $w_\rho$ and $w_z$, we find
\begin{eqnarray}
E_{\rm
int}=\frac{2\pi\hbar^2N^2}{(2\pi)^{3/2}Mw_\rho^2w_z}
\left[a_{\rm ss}
-\frac{\sqrt{5}}{2}a_{\rm sd}\chi(\kappa)\right],
\end{eqnarray}
indeed being real (see also Ref. \cite{andrei2}).
$\kappa\equiv w_\rho/w_z$ is the condensate aspect ratio and
$\chi(\kappa)=\chi_0(\kappa)+\chi_1(\kappa)$ with
\begin{eqnarray}
\chi_0(\kappa)&=&\frac{2\kappa^2+1-3\kappa^2H(\kappa)}{\kappa^2-1},\nonumber\\
\chi_1(\kappa)&=&\frac{6}{5}(\kappa^2-1)H(\kappa),
\end{eqnarray}
and $H(\kappa)\equiv
\tanh^{-1}\sqrt{1-\kappa^2}/\sqrt{1-\kappa^2}$.
The $\chi_0(\kappa)$ term arises from the
bare dipolar interaction Eq. (\ref{redd}) and is known before \cite{yi2,yi3},
while the $\chi_1(\kappa)$ term is due to the momentum-dependent
second term of ${\cal F}_{DD}(\vec k,\vec k')$.

\begin{figure}
\centering
\includegraphics[width=2.8in]{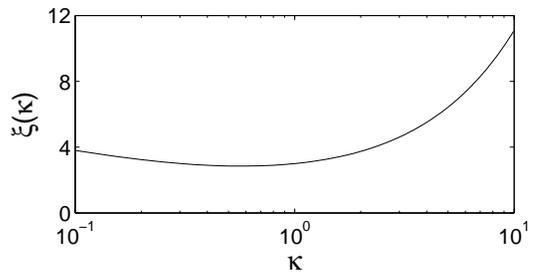}
\caption{The function $\xi(\kappa)$.} \label{fig1}
\end{figure}

Both $\chi_0(\kappa)$ and $\chi_1(\kappa)$ are monotonically
increasing functions of $\kappa$ and vanish at $\kappa=1$. More
specifically, $\chi_0$ is bounded between $-1$ and $2$, while
$\chi_1$ diverges to $-\infty$ and $\infty$ at $\kappa=0$ and
$\infty$. Thus the net dipolar effects from the
pseudo-potential Eq. (\ref{psedd}) are larger, or more prominent
than realized before using $\hat{\cal V}_{\rm DD}^{\rm Born}(\vec R)$.
As a comparison we plot in Fig. \ref{fig1}
$\xi(\kappa)\equiv\chi_1(\kappa)/\chi_{0}(\kappa)$. We note
$\xi(\kappa)$ is relatively flat ($\sim 4$) for
a condensate with moderate aspect ratio $\kappa\simeq 1$, i.e.
the pseudo-potential Eq. (\ref{psedd}) differs from Eq.
(\ref{redd}) only by a scaling factor.
Therefore the pseudo-potential $\hat{\cal V}_{\rm DD}^{\rm Born}(\vec R)$
remains valid near $\kappa\simeq 1$, provided $C_3$ is proportionally scaled
(renormalized) by a factor of $\sim 5$. This enhancement clearly
points to the effect of off-shell $k\neq k'$ collisions. For
on-shell collisions [$k=k'$ in ${\cal F}_{DD}(\vec k,\vec k')$],
approximately a factor of 2 enhancement arises due to the
symmetrization of the scattering amplitude. In the extreme limits
of $\kappa$, collisions are restricted to either 1D or 2D where
different scattering behavior may arise \cite{maxim}, making the
use of Eq. (\ref{psedd}) questionable because $a_{\rm sd}$ is
obtained from the zero energy ($k\rightarrow0$) 3D T-matrix
\cite{mircea}.

Using $a_\rho\equiv\sqrt{\hbar/M\omega_\rho}$ ($\hbar\omega_\rho$)
as unit for length (energy), the dimensionless form of energy
per atom becomes
\begin{eqnarray}
{\cal
E}(q_\rho,q_z)&=&\frac{1}{4}\left(\frac{2}{q_\rho^2}+\frac{1}{q_z^2}\right)
+\frac{1}{4}(2q_\rho^2+\lambda^2q_z^2)\nonumber\\
&&+\frac{{\cal P}_{\rm ss}}{2q_\rho^2q_z}+\frac{{\cal
P}_{\rm sd}}{2q_\rho^2q_z}\chi(\kappa),
\end{eqnarray}
where $q_{\rho,z}=w_{\rho,z}/a_\rho$.
The contact and dipolar interaction parameters are
${\cal
P}_{\rm ss}=\sqrt{2/\pi}\,Na_{\rm ss}/a_\rho$ and ${\cal
P}_{\rm sd}=-\sqrt{5/(2\pi)}\,Na_{\rm sd}/a_\rho$.
The condensate widths
$(q_{\rho0},q_{z0})$ are obtained through a minimization
according to
\begin{eqnarray}
\left.\frac{\partial{\cal E}}{\partial
q_\rho}\right|_{q_\rho=q_{\rho0},q_z=q_{z0}}=\left.\frac{\partial{\cal
E}}{\partial q_z}\right|_{q_\rho=q_{\rho0},q_z=q_{z0}}=0,\nonumber
\end{eqnarray}
which yield
\begin{eqnarray}
q_{\rho 0}&=&\frac{1}{q_{\rho 0}^3}+\frac{{\cal P}_{\rm ss}}{q_{\rho
0}^3q_{z0}}+\frac{{\cal P}_{\rm sd}}{q_{\rho 0}^3q_{z0}}
f(\kappa_0),\label{equi1}\\
\lambda^2q_{z0}&=&\frac{1}{q_{z0}^3}+\frac{{\cal P}_{\rm ss}}{q_{\rho
0}^2q_{z0}^2}+\frac{{\cal P}_{\rm sd}}{q_{\rho 0}^2q_{z0}^2}
g(\kappa_0),\label{equi2}
\end{eqnarray}
where $\kappa_0=q_{\rho0}/q_{z0}$ and
\begin{eqnarray}
f(\kappa)&=&\frac{1}{2(\kappa^2-1)^2}\left[ 4\kappa^4+7\kappa^2-2
-9\kappa^4H(\kappa)\right]\nonumber\\
&&-\frac{3}{5}\left[1-(\kappa^2-2)H(\kappa)\right],\nonumber\\
g(\kappa)&=&\frac{1}{(\kappa^2-1)^2}\left[ 2\kappa^4-10\kappa^2-1
+9\kappa^2H(\kappa)\right]\nonumber\\
&&+\frac{6}{5}\left[1+(2\kappa^2-1)H(\kappa)\right].\nonumber
\end{eqnarray}
A solution $(q_{\rho0},q_{z0})$ is stable if
\begin{eqnarray}
\left[\frac{\partial^2{\cal E}}{\partial
q_\rho^2}\frac{\partial^2{\cal E}}{\partial
q_z^2}-\left(\frac{\partial^2{\cal E}}{\partial q_\rho\partial
q_z}\right)^2\right]_{q_\rho=q_{\rho0},q_z=q_{z0}}>0.
\end{eqnarray}

\begin{figure}
\centering
\includegraphics[width=2.75in]{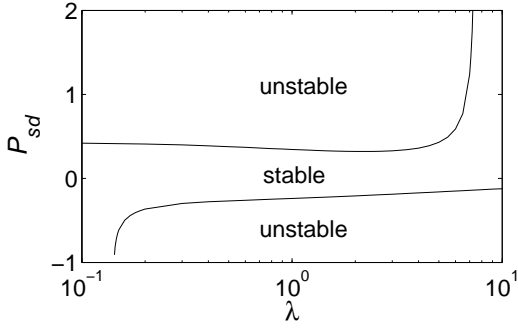}
\caption{The stability diagram of a dipolar condensate when ${\cal
P}_{\rm ss}=0$.} \label{fig2}
\end{figure}

\begin{figure}
\centering
\includegraphics[width=3.25in]{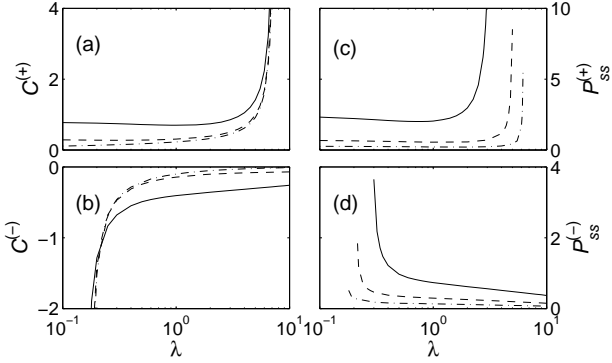}
\caption{The stability diagram for a dipolar condensate with
$a_{\rm ss}>0$. (a) and (b): the $\lambda$ dependence of ${\cal
C}^{(\pm)}$ for ${\cal P}_{\rm ss}=1$ (solid line), $10$ (dashed
line), and $\infty$ (dash-dotted line); (c) and (d): the $\lambda$
dependence of ${\cal P}_{\rm ss}^{(+)}$ (c) and ${\cal P}_{\rm
ss}^{(-)}$ (d) for ${\cal C}=\pm 0.5$ (solid line), $\pm 1$
(dashed line), and $\pm 2$ (dash-dotted line).}\label{fig3}
\end{figure}

We note
$\lambda$ is tunable as mentioned before, one
can also vary $a_{\rm ss}$ with
a Feshbach resonance \cite{stoof,cornish}. Close to
dipolar induced shape resonances, $a_{\rm sd}$ can be
similarly tuned to large or small and positive or negative
\cite{mircea}. The positive valued $a_{\rm sd}$ can be easily
understood by considering the interaction between two polarized
dipoles. If they were placed in a plane perpendicular to their
polarization ($\uparrow\uparrow$), they attract each other; while
they repel each other when placed along the direction of their
polarization ($\rightarrow\rightarrow$), (note the difference
with respect to the bare dipolar interaction \cite{yi1}).
Based on these
considerations, we proceed to study the properties of a dipolar
condensate by varying ${\cal P}_{\rm ss}$, ${\cal P}_{\rm sd}$,
and $\lambda$.

Figure \ref{fig2} shows the stability diagram of a dipolar
condensate when $a_{\rm ss}=0$. For $a_{\rm sd}<0$, the condensate
is stable if ${\cal P}_{\rm sd}<{\cal P}_{\rm sd}^{(+)}$ and there
exists an `always stable' region if $\lambda>\lambda_+\simeq7.56$,
which is greater than the previous estimate \cite{yi2,santos}.
Interestingly, for $a_{\rm sd}>0$, the condensate is stable if
${\cal P}_{\rm sd}>{\cal P}_{\rm sd}^{(-)}$ and it is always
stable for $\lambda<\lambda_-\simeq0.142$.

When $a_{\rm ss}>0$, it is convenient to define ${\cal C}\equiv
{\cal P}_{\rm sd}/{\cal P}_{\rm ss}$ which measures the relative
strength of the dipolar interaction. $\cal C$ can be changed by
tuning $a_{\rm sd}$. As shown in Fig. \ref{fig3} (a) and (b), for
a given $a_{\rm ss}>0$, the condensate is stable if ${\cal
C}^{(-)}<{\cal C}<{\cal C}^{(+)}$. Similar to the previous result
\cite{yi2}, the critical value of $\lambda$ for the `always
stable' region is approximately $\cal C$-independent and equal to
$\lambda_\pm$ for $a_{\rm ss}=0$. For a fixed $\cal C$, the
condensate stability can be modified by tuning ${\cal P}_{\rm
ss}$. From Fig. \ref{fig3} (c) and (d), we see that only when
${\cal P}_{\rm ss}<{\cal P}_{\rm ss}^{(+)}$ for ${\cal C}>0$ and
${\cal P}_{\rm ss}<{\cal P}_{\rm ss}^{(-)}$ for ${\cal C}<0$ is
the condensate stable. Varying ${\cal P}_{\rm ss}$ for a constant
$\cal C$ can be achieved by changing $N$.

\begin{figure}
\centering
\includegraphics[width=2.8in]{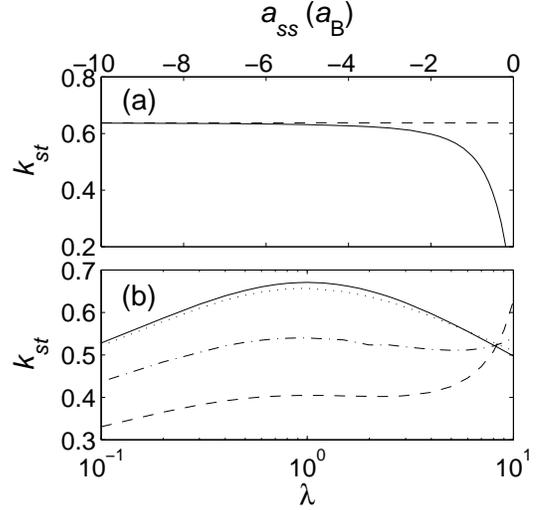}
\caption{(a) The dependence of $k_{\rm st}$ on $a_{\rm ss}$ for
$\lambda=6.8/17.35$. The dashed line is the result without dipolar
interaction. (b) The dependence of $k_{\rm st}$ on $\lambda$ for
$a_{\rm ss}=-0.5(a_B)$ (dashed line), $-1(a_B)$ (dash-dotted
line), and $-4(a_B)$ (dotted line). The solid line is the result
without dipolar interaction.} \label{fig4}
\end{figure}
\begin{figure*}
\centering
\includegraphics[width=6.8in]{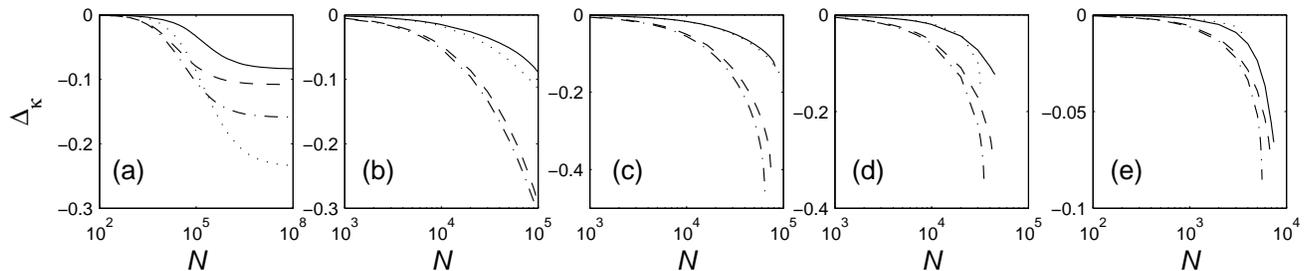}% Here is how to import EPS art
\caption{The $N$ dependence of $\Delta_{k}$ for $a_{\rm
ss}=5(a_B)$ (a), $0.5$ (b), $0$ (c), $-0.5$ (d), $-5$ (e), and
$\lambda=0.2$ (solid line), $0.6$ (dashed line), $2$ (dash-dotted
line), $6$ (dotted line).}\label{fig5}
\end{figure*}

When $a_{\rm ss}<0$, the
condensate becomes unstable if $N$ exceeds a critical value
$N_{\rm cr}$ even without dipolar interaction.
This instability is typically measured by the
stability coefficient $k_{\rm st}\equiv N_{\rm cr}|a_{\rm
sc}|/a_{\rm ho}$ \cite{ane,garmmal} with $a_{\rm
ho}=\sqrt{\hbar/m\bar\omega}$ and
$\bar\omega=(\omega_x\omega_y\omega_z)^{1/3}$. To study the
dipolar induced modification of $k_{\rm st}$, we consider the
experiment \cite{donley,cornish} of $^{85}$Rb atoms in
state $|F=2,M_F=2\rangle$ with a magnetic dipole moment of
$\mu=2\mu_B/3$. We take $a_{\rm sd}$ as obtained for the
multi-channel scattering calculation \cite{mircea}.
For $\omega_\rho=(2\pi)17.35$ (Hz), we then
have ${\cal P}_{\rm sd}\approx 5.0\times 10^{-6}N$. Figure
\ref{fig4} (a) shows the $a_{\rm ss}$ dependence of $k_{\rm st}$
for $\lambda=6.8/17.35$. As before
 dipolar interaction destabilizes a
condensate for this configuration; However, the enhanced effect
due to the more rigorous pseudo-potential Eq. (\ref{psedd}) is
more profound: even for $a_{\rm ss}<-3(a_B)$, the effect of
dipolar interaction is still visible. In Fig. \ref{fig4} (b), we
plot the $\lambda$ dependence of $k_{\rm st}$. Since ${\cal
P}_{\rm sd}>0$, dipolar interaction destabilizes the condensate at
small values of $\lambda$; when $\lambda$ is large, dipolar
interaction can also stabilize an otherwise attractive condensate.

Finally, we briefly consider
the condensate aspect ratio $\kappa_0$.
We define its relative change \cite{yi3}
\begin{eqnarray}
\Delta_{\kappa}\equiv\frac{\kappa_0({\cal P}_{\rm sd}\neq
0)-\kappa_0({\cal P}_{\rm sd}=0)}{\kappa_0({\cal P}_{\rm sd}=0)}
\end{eqnarray}
as a measure of whether it is possible to detect the dipolar
interaction from imaging condensate shape within current
experiments. Figure \ref{fig5} shows its $N$-dependence for
various $a_{\rm ss}$ and $\lambda$. For certain $\lambda$,
$\Delta_{\kappa}$ can be as high as 20\% even with
$N=2\times10^5$, if $|a_{\rm ss}|$ is tuned small.

In conclusion, we have calibrated the properties of a
dipolar condensate using the more rigorous anisotropic
pseudo-potential \cite{andrei1}, based on a variational calculation.
Significant enhancement were found to all dipolar effects
predicted previously. In the limit of weak s-wave interactions,
when $P_{\rm ss}\le 1$ due to a small $a_{\rm ss}$ and/or a small $N$,
our results are clearly valid based on previous studies
with the same variational technique \cite{yi2,yi3,yi4,var,var1},
where extensively comparisons were preformed to justify
the trial function Eq. (\ref{trial}) \cite{var,var1},
including the presence of a weak dipolar interaction \cite{yi2,yi3,yi4}.
We note this is also the interesting
limit where experimental detection of coherent dipolar interactions
will likely occur.
The variation approach gives reliable
stationary condensate properties, although not the density profile itself,
even when the interaction energy is large \cite{var1}.
The results from $P_{\rm ss}>1$, although consistent, needs
further improvements with more accurate numerical calculations
(over the present variational method).

We acknowledge insightful discussions and communications
with Dr. Derevianko. This work is supported by NSF and NSFC.

\end{document}